# Willingness to pay, surplus and Insurance policy under dual theory

## Neji SAIDI [1]

## Abstract

In this paper, we aims to state some proprieties of willingness to pay (WTP) for partial risk reduction and links with insurance within the dual theory of decision. In the case of partial reduction, we get as Langlais (2005) that a risk-averse decision maker (DM) can have a willingness to pay small than this of a neutral one. By decomposition the WTP as Courbage and al (2008), we get that a strong averse DM is willing to give more for a reduction of a high probability portion rather than a low probability one.

The main result is that in the dual theory, reducing probability of risk and supply insurance can be complementary if the surplus is increasing in risk reduction.

## JEL Classification: C61, D81, D82 & G22

---

[1] Assistant Professor, *College of Sciences, King Faisal university, KSA and Higher Institute of Computer Sciences of Kef, University of Jendouba, Tunisia*
*E-mail: nsaidi2010@gmail.com or nsaidi@kfu.edu.sa . tel + 966 557535221.*
*Address: PO: Math-department , King University Faisal ,KSA*

# 1. Introduction

Risk management can take many forms. A decision maker (DM) can transfer risk to an insurance company against a premium. Similarly, it may invest in self-protection activities to reduce the risk.

Several researchers have studied the relationship between insurance and risk reduction. Kahenmann and Tversky (1979) were the first to compare insurance and "probabilistic insurance". In the first situation, the holder disposes of transferring risk to an insurance company. However, in the second case, a third party (government, insurance, ...) acts to reduce the probability of risk against part of a sum called willingness to pay or WTP .

According to Ehrlich and Becker (1972), DM attempts to reduce the likelihood of a disaster by self-protection. In addition, risk reduction can be public action such as improving roads states to reduce the risk of accident or making vaccinations to avoid contamination.

Risk reduction activity can be advantageous, especially for undiversified disaster (or where insurance cannot intervene or if people have financial constraints. We can mention the interesting case of such activities is agriculture, where the actions of public or collective protection are beneficial.

Proprieties of WTP have been widely studied (Ehrlich and becker1972; Eeckhoudt and al (1997); Dachraoui and al (2004)). Godfroid (2000) has also focused on the complementarities between reduction risk activities and insurance market.

Courbage (2001) give some proprieties of self-protection and insurance within the dual theory of choice. Also, Bleichrodt and al(2006) have studied willingness to pay in the case of health risks within the RDEU model willingness to pay. However, Langlais (2005) compared proprieties of WTP in the case of first and second stochastic dominance.

Our work sheds lights on models those introduce weightings of probabilities like the rank dependent utility models or RDEU models. Our principal purpose is to analyze some proprieties of willingness to pay and insurance in the dual theory or Yaari model. An important argument why DM use dual theory is that he does not evaluate probabilities linearly, but he introduce probability distortion. This alteration leads to a probability weighting that describe the DM beliefs.

We get that for the range of probabilities the willingness to pay for a risk-averse DM is less than that observed for a risk-neutral DM. In addition, reduction of probability and insurance can be complementary.

Our paper is organized as follows: in section 2, we define willingness to pay for total or partial risk reduction as Langlais (2005). In section 3, we describe briefly the model of Yaari and analyze proprieties of WTP and insurance. In section 4, before conclusion, we study the complementarily of those two types of risk management.

## 2. Willingness to pay for risk reduction

Let a decision maker (DM) endowed with initial wealth $W_0$ and a risky asset L where loss probability is $p_0$. So his wealth is $W=W_{init} = W_0 + X_0$ with: $X_0 =(0, p_0; L, 1-p_0)$.

Decision maker choices are characterized by the preference relation $\underset{\sim}{\phi}$. This relation was supposed to be a total preorder on the set of lotteries. We denote $\delta_b$ the degenerate random giving the result $b$ certainly.

**Definition 1**

The risk premium $\pi(W_0, p_0)$ is given by:

$$W_0 + X_0 \sim \delta_{W_0+E(X_0)-\pi(W_0,p_0)} \quad (1)$$

It represents the maximum amount an individual is willing to give for the expected wealth *E(W)*.

An individual is said to be weakly risk averse if he prefer $E(W)$ to $W$. Then, it is easy to see that his risk premium is strictly positive

**Definition 2**

Willingness to pay or WTP, noted $v(W, p_0, 0)$ to eliminate risk is such that:

$$W_0 + X_0 \sim \delta_{W_0 + L - v(W, p_0, 0)} \quad (2)$$

It is the maximum amount that the agent is willing to give in order to reduce the probability $p_0$ to $0$.

**Definition 3**

Willingness to pay for a partial probability (risk) reduction from $p_0$ to $p_1$ denoted $v(W, p_0, p_1)$, where $p_1 < p_0$, is defined by:

$$W_0 + X_0 \sim W_0 + X_1 - v(W, p_0, p_1) \quad (3)$$

$$\text{where } X_1 = (0, p_1; L, 1 - p_1)$$

**Proposition 1**

The WTP of total reduction for risk averse DM is above that of a risk-neutral DM.

**Proof 1**

Let $v_N(W, p_0, 0)$ the WTP for a neutral DM. It is easy to see that the WTP[2] is none other than: $v(W, p_0, 0) = p_0 L + \pi(W, p_0)$ with $\pi(W, p_0) > 0$

Then: $v(W, p_0, 0) = p_0 L + \pi(W, p_0) > p_0 L = v_N(W, p_0, 0)$.

### 3. Willingness to pay for partial risk reduction in Yaari Model

In Yaari model (1987), the relation $\underset{\sim}{\phi}$ is presented by a function $DT$ defined for any lottery $W = (w_f^1, p_1; w_f^2, p_2; ...... w_f^n, p_n)$ with $w_f^1 < w_f^2 < ...... < w_f^n$ by:

---
[2] Willingness to pay

$$DT(W) = \sum_{k=1}^{k=n} v_k w_f^k, \quad \text{where} \quad v_k = f(\sum_{j=k}^{n} p_j) - f(\sum_{j=k+1}^{n} p_j)$$

In this model, the shape of the transform function of probabilities describe the behavior towards risk.

In this paper, we Assume that the behavior of the agent is adequate to model Yaari. and that the transformation function probability $f$ is class $C^2$, increasing over *[0,1]* with *f(1)=1* and *f(0)=0*. Thus, we agree that it is characterized by a differentiable function $f$ such that it evaluates its initial wealth by:

$$DT(W_{init}) = [1 - f(1-p)]W_0 + f(1-p)(W_0 + L) \qquad (4)$$

*Proposition2*

Relation (4) can be written as:

$$v(W, p_0, p_1) = DT(X_1) - DT(X_2) = v(W_0, p_0, 0) - v(W_0, p_1, 0)$$

The last result is similar to Courbage and al(2008) which indicated that we can compute WTP for partial risk reduction between two states if we now WTP to eliminate risk in each state.

*Proposition3*

If the DM's behavior is consistent with the model of Yaari, then there exists a real c between $p_1$ and $p_0$ that:

$$v(W, p_0, p_1) = f'(c)[p_0 - p_1]L \qquad (5)$$

*Proof*

Indeed $DT(W) = [1 - f(1-p_0)]W_0 + f(1-p_0)(W_0 + L) = W_0 + f(1-p_0)]L$

$= DT(W_0 + X_1 - v(W_0, p_0, p_1)) = W_0 - v(W, p_0, p_1) + f(1-p_1)L$

Then, $f(1-p_0)L = -v(W, p_0, p_1) + f(1-p_1)L$

Since f is differentiable, then simply apply the mean value theorem

### *Proposition 4*

If DM is strongly risk averse, then

a) His WTP for partial risk reduction can be less than the WTP for a neutral one.

b) The WTP for a rate of proportion reduction with a high probability of loss is greater than that of a low probability

### *Proof*

a) In Yaari model, strong averse toward risk is described by a convex function f. Since f' is increasing, then: $v(W, p_0, p_1) < f_1'(p_0)(p_0 - p_1)L$. Therefore, it is sufficient to have $f_1'(p_0) \leq 1$[3] to assure that: $v(W, p_0, p_1) < v_N(W, p_0, p_1) = (p_0 - p_1)L$

b) From 5, if we need to reduce a proportion $\alpha$ from $p$, then there exist $\theta \in \,]0,1[$ ; $v(W, p, (1-\alpha)p) = f'(\theta p)[(1-\alpha)p]L$.

In addition, $\dfrac{\partial v}{\partial p}(W_0, p, (1-\alpha)p_1) = \theta f''(\theta p)[(1-\alpha)p]L + (1-\alpha)f'(\theta p)L > 0$. Which means that DM is willingness to pay more for reducing a proportion for high probabilities of loss, than for the same proportion if probability is low.

This result joined the remarks raised in the model of expected utility where a more risk averse decision maker does not necessarily invests more money in activities reductions probability(Eeckhoudt and al (1997)). Also, Langlais(2005) get similar results in a model without expected utility assumption.

### 4. Willingness to pay in the presence of insurance

The relationship between self-protection or probability reduction and market insurance is ambiguous. Ehrilich and Becker (1972) were the first to report that insurance and auto insurance are substitutes. However, the insurance and self-protection can also be complementary (Courbage (2001)).

---

[3] Bleichrodt and al(2006) said that individual is relatively inentive to chnge in point $p_0$.

Similarly, we analyze the willingness to pay in the presence of insurance in case of probability distortion.

### 4.1. Insurance demand

Suppose the individual seek insurance against loss. if he buy a cover $Q \in [0, L]$ from insurance market where $\lambda$ is the loading rate, then he pay a premium:

$$\pi(Q) = p_0 Q(1 + \lambda) \qquad (6)$$

**Proposition 5**

(i) DM will accept insurance if $\lambda \leq \dfrac{(1 - p_0) - f(1 - p_0)}{p_0} = \lambda^* \qquad (7)$

(ii) If loading rate is below $\lambda^*$, DM utility will be maximal with full insurance.

*Proof proposition 5*

i) The DM accept insurance if:

$$DL(W_{final}) = [(1 - p_0(1 + \lambda)) - f(1 - p_0)]Q + [W_0 + f(1 - p_0)L] \geq DT(W)$$

This leads to $[(1 - p_0(1 + \lambda)) - f(1 - p_0)] \geq 0$ or: $1 + \lambda \leq \dfrac{1 - f(1 - p_0)}{p_0}$

ii) $\underset{Q \leq L}{Max} \; DL(W) = [(1 - p_0(1 + \lambda)) - f(1 - p_0)]Q + [W_0 + f(1 - p_0)L]$

$$= a(\lambda)Q + [W_0 + f(1 - p_0)L]$$

Or $a(\lambda)Q > 0$, then $Q_{max} = L$

The first part of proposition emphasizes the importance of beliefs to accept insurance. In addition, the second one shows that insurance decision is a corner solution[4].

---

[4] See Doherty et al (1995)

(i) $\mathrm{DT}(W)\Big|_{Q^*\leq L} = \left[(1-p_0(1+\lambda)) - f(1-p_0)\right]Q + \left[W_0 + f\,(1-p_0)L\right] \geq DT(W_{init})$

$DT(W)\Big|_{Q^*\leq L} = ((1-p_0(1+\lambda)) - f(1-p_0))Q + [W_0 + f(1-p_0)L]$

$DT(W)\Big|_{Q^*\leq L} \geq DT(W_{init})$ if $1 - f(1-p_0) \geq p_0(1+\lambda)$

(ii) $MaxDT(W)\Big|_{Q\leq L} = \left[(1-p_0(1+\lambda)) - f(1-p_0)\right]Q + \left[W_0 + f\,(1-p_0)L\right]$

$= a(\lambda)Q + [W_0 + f\,(1-p_0)L];\quad a(\lambda) > 0$

Is obtained if $Q^* = L$

## 4.2. Insurance policy *and willingness to pay*

The agent choose insurance if the coverage improve his utility. Else, he will refuse any insurance contract. Proposition 5 indicates that agent will manage risk by coverage if the loading rate is below the threshold $\lambda^*$ which depends in part on the probability $p_0$.

Upcoming, we will suppose that: $1 + \lambda \geq \dfrac{1 - f(1-p_0)}{p_0}$ or $\lambda > \lambda^*$, which implies that the agent will refuse to take insurance because his utility will decrease. Therefore, the insurer seek for technology to reduce loss probability. He will beer a cost c(x) for any reduction x.

Our work differs from Courbage (2001) because probability reduction here is an insurer policy but is not a self-protection.

### 4.2.1. The model

The insurer, which we suppose risk neutral, can seek to reduce risk probability if the agent is able to pay the cost. He will bear a loss of the prime insurance if he accept to reduce probability from $p_0$ to $p_0-x$. In the other hand, he will receive the amount of the risk reduction equal to $TC(x)=c(x).L$. Where $c(x)$ denotes the unity cost.

Next, we will note $WTP(W, p_0, x) = v(W, p_0, p_0 - x)$

Therefore, The insurer policy can be achieved if $c(x)L \leq WTP(W, p_0, x)$.

**We denotes** $S(p_0, x) = WTP(W, p_0, x) - TC(x) \geq 0$ **the surplus or benefit of probability reduction from $p_0$ by x portion.**

Let $E_f = \{0 < x < p_0 / TC(x) \leq WTP(W, p_0, x)\}$ the set of admissible actions of the insurer.

**We assume that $E_f$ is non empty.**

The insurer optimization problem is:

$$\underset{x \in E_\lambda}{Max} (p-x)(1+\lambda)\alpha(x)L + c(x).L$$

(8)

**s/c:** $W_0 - c(x).L + \alpha L - (p-x)(1+\lambda)\alpha L) + f(1-(p-x)(1-\alpha)L \geq W_0 + f(1-p)L$ **(PC)**

**Or** $WTP(P, x) - c(x).L \geq \alpha L[(p-x)(1+\lambda) - 1 + f(1-(p-x))]$

That means, the amount of proportion assured is less than the willingness to pay for reduce the probability p by x percent.

We can see that for neutral agent with fair insurance ($\lambda = 0$), the (PC) constraint is reduced to $WTP(P, x) - c(x) \geq 0$ or $x \in E_f$.

4.2.2. **Discussion**

a) First case: if x=0, then (PC) is not verified. Hence, agent will not sell any coverage

b) Second case: 0<x<p: reduction risk policy

The program will be:

$$\underset{(x,\mu)}{Max} \Pi =$$

$$(p-x)(1+\lambda)\alpha(x)L+c(x)L+\mu[\alpha-(p-x)(1+\lambda)\alpha L+f(1-(p-x)(1-\alpha)L-f(1-p)L-c(x)L] \quad (9)$$

The optimal level of risk reduction verify:

$$\frac{\partial \Pi}{\partial x}=-(1+\lambda)\alpha L+(p-x)(1+\lambda)\alpha'(x)L-c'(x)L+\mu[\alpha'(x)+(1+\lambda)\alpha L-f'(1-(p-x)(1-\alpha)L-c'(x)L]=0$$

The optimal impact of the constraint is :

$$\frac{\partial \pi}{\partial \mu}=\alpha L-(p-x)(1+\lambda)\alpha L+f(1-(p-x)(1-\alpha)L-f(1-p)L-c(x)L$$

We look the case where the insurer like to extract all the surplus from the agent, then the (PC) will be satured and $\mu > 0$. Then:

$$\alpha_\lambda(x)=\frac{f[1-(p-x)]-f(1-p)-c(x)}{[(p-x)(1+\lambda)+f[1-(p-x)]-1]} \quad (10)$$

$$\alpha_\lambda(x)=\frac{WTP(p,x)-TC(x)}{WTP(p,x)+A(x)} \quad (11)$$

Where $A(x)=[(p-x)(1+\lambda)+f(1-p)-1]L$

Let us define $\frac{\partial}{\partial x}[WTP(p,x)-TC(x)]$ the marginal surplus from reduction of x portion.

**Proposition 6** : if $0<x<p-\frac{1-f(1-p)}{1+\lambda}$ and the marginal profit of risk reduction is positif, then :

a) $\alpha_\lambda(x)$ is increasing in x.

b) Coverage is partial

The portion $\alpha_\lambda(x)$ increasing in x indicate *insurance and the probability reduction are complementarians. Also,* coverage partial differe from the case of corner solution in standard choice of insurance in Yaari model

## Proof proposition 6

a) $\dfrac{\partial \alpha}{\partial x} = \dfrac{[f'(1-(p-x))-c'(x)][(p-x)(1+\lambda)+f(1-p)-1]+(1+\lambda)[f[1-(p-x)-f(1-p)]]}{[(p-x)(1+\lambda)+f[1-(p-x)]-1]^2}$

$\dfrac{\partial \alpha}{\partial x} = \dfrac{[f'(1-(p-x))-c'(x)]A(x)+(1+\lambda)WTP(p,x)}{L.[(p-x)(1+\lambda)+f[1-(p-x)]-1]^2}$

**Or** $0 < x < p - \dfrac{1-f(1-p)}{1+\lambda}$, that means $[(p-x)(1+\lambda)+f(1-p)]-1 > 0$ or also

A(x)>0. Which leads to $\dfrac{\partial \alpha}{\partial x} > 0$

b) If $0 < x < p - \dfrac{1-f(1-p)}{1+\lambda}$ then $WTP(p,x) - TC(x) < WTP(p,x) + A(x)$.

Therefore : $\alpha_\lambda(x) < 1$.

**Interpretation: if insurer can reduce risk of contamination by a cost**

### 5. Conclusion

This paper is the first step of WTP and insurance analysis in models including people beliefs. Our paper is restricted to the binary case and we have restricted our analysis to the Yaari model. This work can be extended to more general model like the RDEU framework.